\documentclass[a4paper,amsmath,prl,onecolumn]{revtex4-1}
\usepackage{graphicx}
\usepackage{bm,color}
\newcommand{\white}[1]{{\color{white}{#1}}}
\begin{document}
\title{Early experimental evidence of a topological quantum state\\ The signature of the Haldane ground state revealed by scattered neutrons}
\author{M Enderle}
   \affiliation{Institut Laue-Langevin, CS 20156, F-38042 Grenoble C\'{e}dex 9, France}
\author{M Kenzelmann}
  \affiliation{Paul-Scherrer Institut, CH-5232 Villigen PSI, Switzerland}
\author{WJL Buyers}
  \affiliation{Canadian Neutron Beam Centre, Canadian Nuclear Laboratories, Chalk River, Ontario K0J~1J0, Canada}

\begin{abstract}
Neutron scattering played an important role in the experimental exploration of the theoretical ideas of Thouless, Haldane, and Kosterlitz, who won the Nobel Prize in Physics in 2016. This article reviews the signatures of Haldane's predicted topological quantum ground state and their observation in early neutron scattering experiments, which overturned the wisdom of the day.
\end{abstract} 

\maketitle

The quantum behaviour of assemblies of many interacting particles leads to surprising and counter-intuitive new phenomena: superconductivity and superfluidity are well-known examples. Duncan Haldane discovered a new quantum mechanical phenomenon \cite{1}: one-dimensional antiferromagnets ({\it spin chains}) have an entirely different quantum ground state depending on whether the individual spins are integers (1,2,3, \ldots) or half-odd integers (1/2, 3/2, \ldots). This fundamental distinction is inherently related to the fact that one turn brings an integer spin back to its initial state, while a half-odd integer spin needs two turns. 

Using this peculiar trait of quantum mechanics, Haldane made precise predictions that could directly be tested by neutron scattering experiments (see cartoon): integer spin chains have a cooperative singlet ground state with only short-range pair correlations. An energy gap (very large for spin 1, see cartoon) separates the ground state from the elementary excitation triplet. In contrast, the singlet ground state of half-odd integer spin chains features far-reaching spin pair correlations that decay algebraically and the excitation spectrum is gapless. The half-odd integer spin ground state is therefore fragile with respect to long-range magnetic order in the presence of interchain interactions whereas the integer spin chain ground state is protected by the energy gap. Hence a spin-1/2 antiferromagnetic chain seems closer to the {\it classical} antiferromagnetic N\'{e}el long-range order than the spin-1 chain. Since quantum effects scale with 1/S, it was generally expected that there is an increasing tendency towards long-range order with increasing spin value. While theorists at the time were working hard to elaborate this puzzle, the new idea was immediately tested experimentally, and early neutron experiments provided ''proof'' prior to a full theoretical understanding.

In real materials, excitation gaps arise for a variety of reasons, amongst which anisotropy and dimerisation were well-known at the time. The first evidence for the {\it Haldane gap} was found in the quasi-one dimensional spin-1 compound CsNiCl$_3$ for temperatures above its N\'{e}el temperature \cite{2}, where magnetic order is absent. The absence of quasi-elastic scattering provided evidence for a singlet ground state, and the observed excitation gap was much larger than what spin anisotropies derived from linear spin-wave theory at low temperature could explain. Polarised neutron experiments provided the first direct proof of the isotropic triplet character of the gapped excitation in CsNiCl$_3$ \cite{3} and excluded anisotropy as origin of the gap. Polarised neutron data also proved Zeeman splitting of the triplet \cite{4} and triplet character along the entire one-dimensional dispersion \cite{5}. For an early theoretical review with a short summary of the early experiments, see e.g. Ref.~\cite{6}.

The first spin-1 chain discovered that did not order at any temperature was NENP (Ni(C$_2$H$_8$N$_2$)$_2$NO$_2$(ClO$_4$)) \cite{7}. Initially, the origin of the excitation gap in NENP was not clear due to the large anisotropy, and dimerisation could not be excluded either. The Zeeman splitting of the three gapped excitations, and the precise ratio between anisotropy and exchange definitely ruled out a trivial anisotropy gap \cite{8,9}. The dispersion of both CsNiCl$_3$ \cite{10} and NENP \cite{9} was found to have a periodicity of $2\pi/d$, reflecting the unbroken translation symmetry by one spin spacing $d$ of the Haldane ground state. In both compounds the elementary triplet excitations were shown to dominate the spectral weight \cite{8,9,11}.

Two spin chain materials of identical crystallographic structure, AgCrP$_2$S$_6$ and AgVP$_2$S$_6$  with spin 3/2 (Cr) and spin 1 (V), respectively, allowed the first direct comparison of half-odd integer and integer spin chains \cite{12}. The spin 3/2 compound showed long-range order and gapless excitations, while long-range order was absent and the excitation spectrum gapped for the isostructural spin-1 compound. 
Quasi-elastic studies of the correlation length as function of temperature in a series of isostructural ABX$_3$ compounds resulted in a finite correlation length towards zero temperature for the spin-1 compounds, while the half-odd integer compounds tended to infinitely long-ranged correlations \cite{13}. The nature of the one dimensional quantum ground state strongly influences even the long-range ordered phases of weakly coupled spin chains: CsMnI$_3$ (S=5/2), isostructural to CsNiCl$_3$, has standard spin-wave excitations and field dependence \cite{14}, in contrast to CsNiCl$_3$ where the spectrum in the long-range ordered phase features renormalized {\it Haldane triplets} that interact in adjacent chains \cite{15}. 

Duncan Haldane also pointed out \cite{1} that the one-dimensional quantum spin-1 antiferromagnet at zero temperature is related to the classical two-dimensional planar ferromagnet at finite temperature. According to Kosterlitz and Thouless, the latter has vortex-like topological excitations ({\it skyrmions}) that condense into the ground state upon lowering the temperature. The ground state of the quantum spin-1 chain is characterised by a {\it topological string} order \cite{16}, which is analogous to the topological order responsible for the fractional quantum Hall effect \cite{17} and sketched in the cartoon.

\newpage

\begin{center}
{\large Cartoon: The Haldane ground state in a nutshell (isotropic case)}
\end{center}

\setlength{\parindent}{0pt}

\begin{tabular}{p{12.5cm}p{4cm}}
\hspace{-0.5em}\begin{minipage}[b]{12.5cm}
Consider isotropic magnetic interactions between nearest-neighbour spins that favour antiparallel alignment independent of the spin direction in space:
\end{minipage} & \includegraphics[width=4cm]{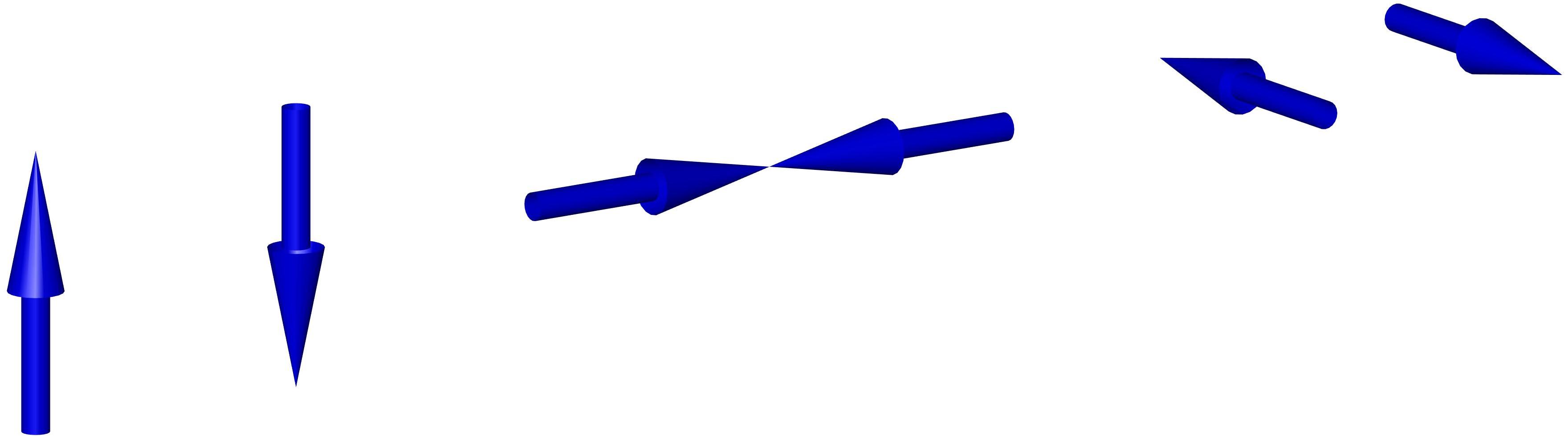}
\end{tabular}

\bigskip

At $T=0$, an isotropic classical (i.e. $S\to \infty$) one-dimensional antiferromagnet is in the N\'{e}el state with spontaneously broken translation symmetry:

\begin{center}
\includegraphics[width=0.6\textwidth]{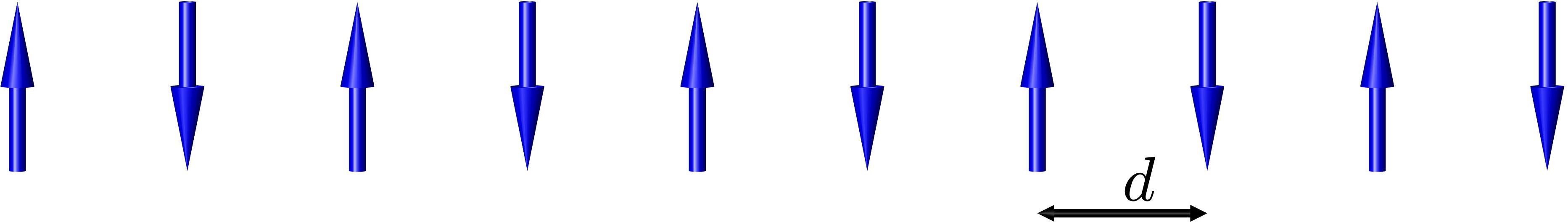}
\end{center}

\begin{tabular}{p{3cm}p{8.5cm}p{5cm}}
\hspace{-0.5em}\begin{minipage}[b]{3cm}
This strictly periodic pattern of spin states leads to sharp Bragg peaks 
in neutron diffraction. \\
\white{.}\\
\white{.} \\
\vfill
\end{minipage} & \includegraphics[width=8.5cm]{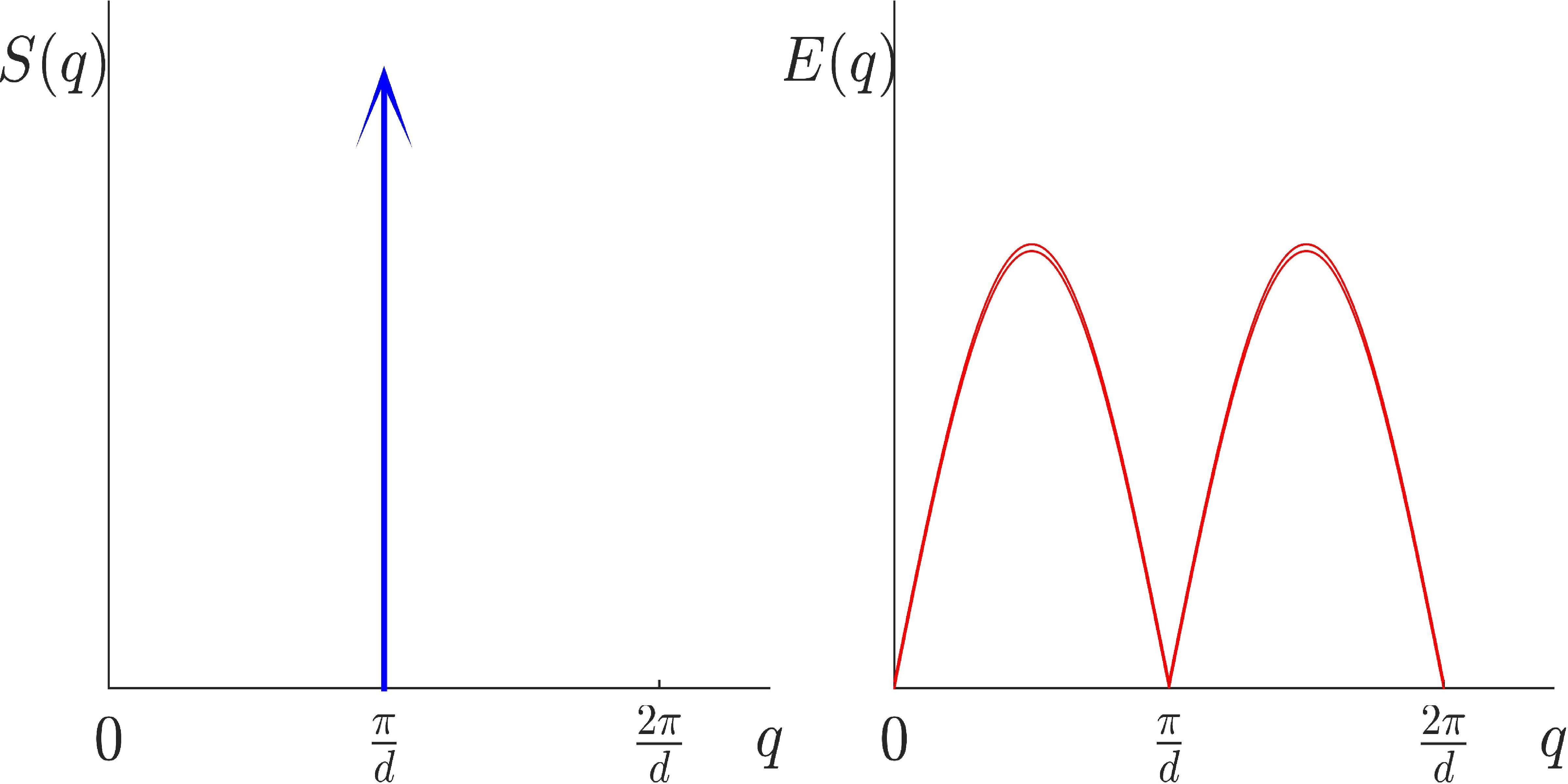} & \begin{minipage}[b]{5cm}
The excitations are doubly degenerate spin waves, small deviations from the ordering direction. Their energy-momentum relation (dispersion) is gapless and has period $\pi/d$ in $q$ (broken translation symmetry). \\
\white{.} \\
\white{.} \\
\vfill
\end{minipage}
\end{tabular}

\bigskip

Quantum spins 1/2 form a coherent collective singlet ground state with no local order of the spins. This is {\it not} the N\'{e}el state, but a snapshot would still show infinitely large regions that look like the classical N\'{e}el state. The translation symmetry is unbroken. Since there are no ordered moments there are no Bragg peaks,

\begin{tabular}{p{3cm}p{8.5cm}p{5cm}}
\hspace{-0.5em}\begin{minipage}[b]{3cm}
but the peak intensity diverges in a power-law, i.e. with an infinite correlation length like at a critical point.\\ 
\white{.} \\
\vfill
\end{minipage} & \includegraphics[width=8.5cm]{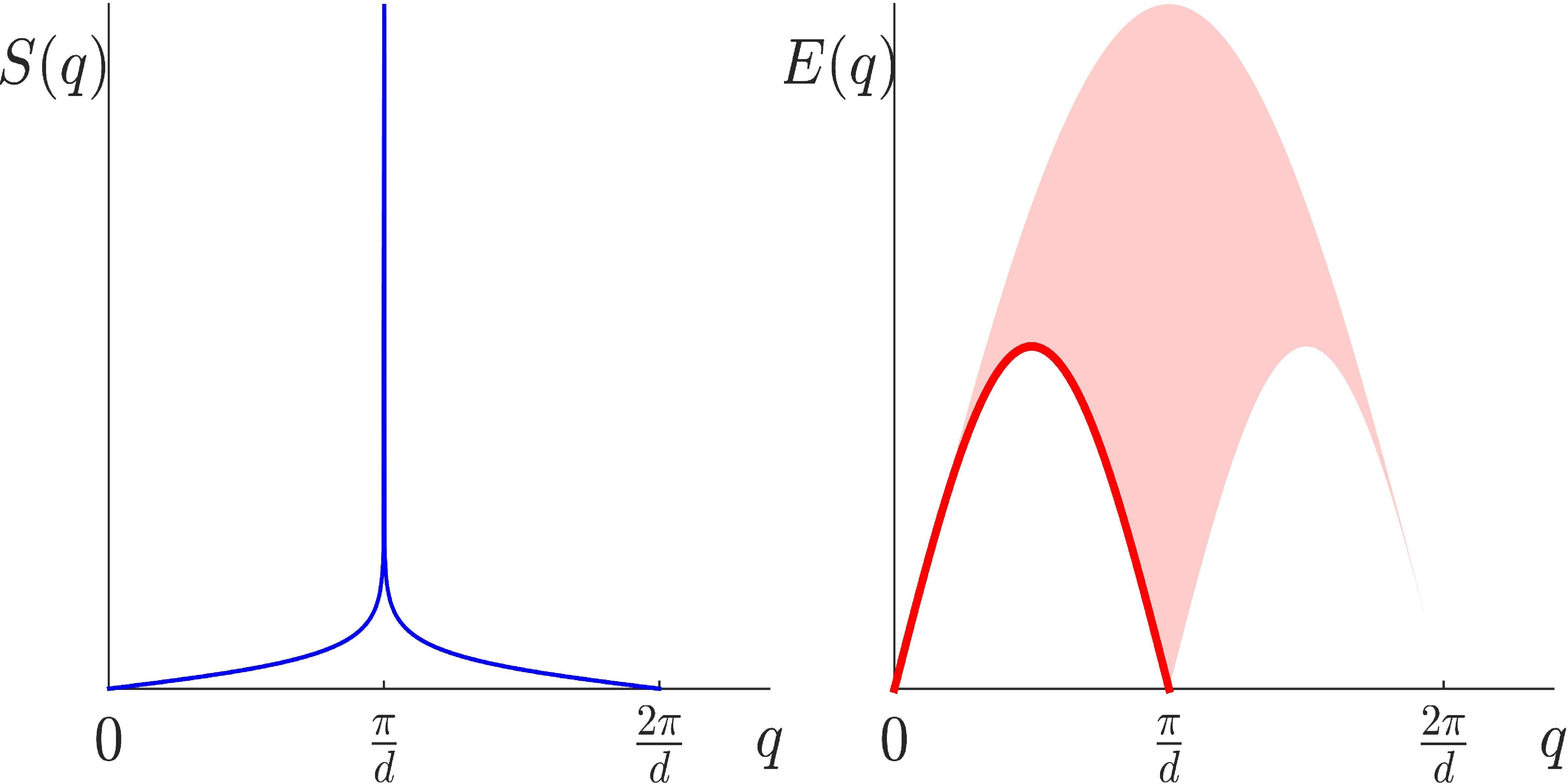} & \begin{minipage}[b]{5cm}
The excitations are pairs of topological (domain-wall like) spinons with a gapless dispersion that repeats with period $2\pi/d$ (unbroken translation symmetry). \\
\white{.} \\
\white{.} \\
\vfill
\end{minipage}
\end{tabular}

\smallskip

Even a tiny interchain coupling leads to long-range order in real spin 1/2 materials.

\bigskip

Haldane \cite{1} predicted that integer spin (1,2,3, \ldots) antiferromagnetic chains appear even less ordered,

\begin{tabular}{p{1.5cm}p{11cm}p{3.5cm}}
\hspace{-0.5em}\begin{minipage}[b]{1.5cm}
with broad non-diverging 
peaks in diffraction. \\ \white{.} \\
\vfill
\end{minipage} & \includegraphics[width=11cm]{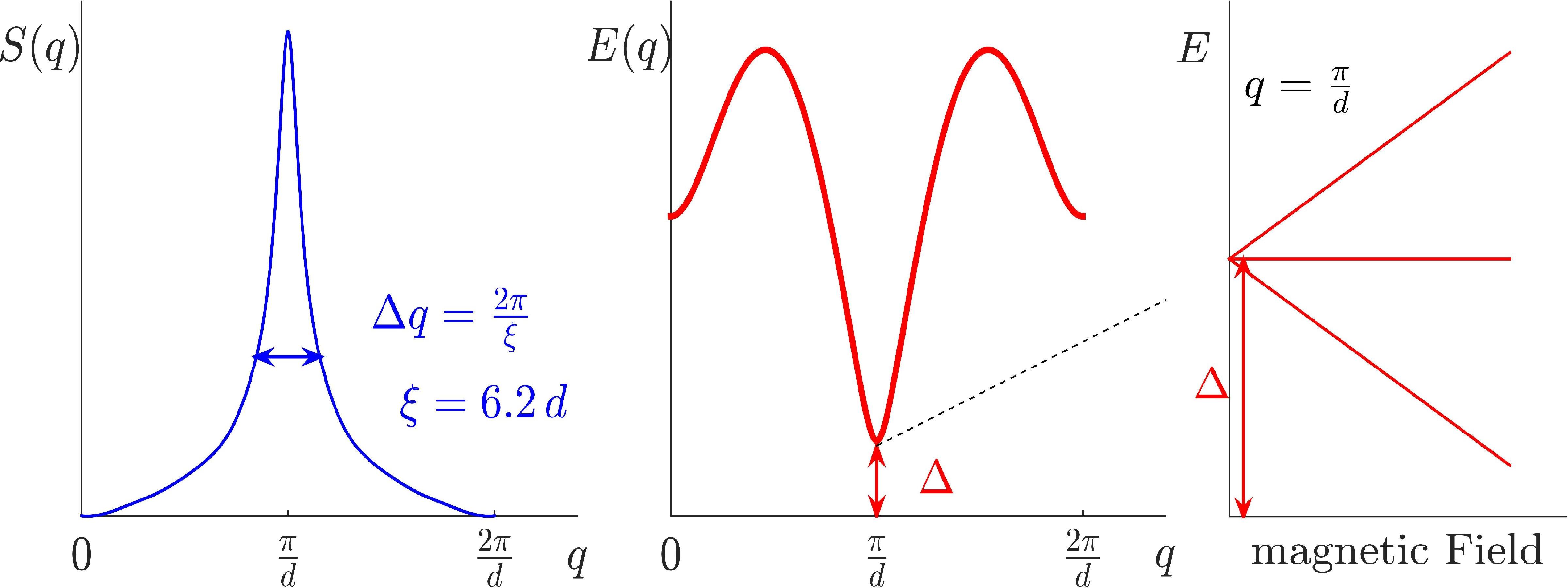} &
\begin{minipage}[b]{4cm}
The elementary excitations have a finite minimum energy ({\it Haldane gap}), are sharp  in energy and triply degenerate. The period  $2\pi/d$ of the dispersion reflects the unbroken translation symmetry. \\
\vfill
\end{minipage}
\end{tabular}

\medskip

Half-odd integer (1/2, 3/2, \ldots) antiferromagnetic spin chains should resemble the gapless spin 1/2 case.

\newpage

For spin 1, the Haldane ground state singlet is characterized by a {\it topological string} (i.e. non-local) order \cite{16}. This string order can be understood as a {\it dilute} antiferromagnetic order: spin up ($S^{z}=1$) needs to be followed by spin down ($S^{z}= - 1$) or any number of diluting zeros ($S^{z}=0$), and spin down by spin up or any number of zeros. Antiferromagnetic correlations remain short-ranged.

\bigskip 

\begin{center}
\begin{tabular}{p{3cm}p{13cm}}
\begin{minipage}[c]{3cm}
ground state\\
\white{.}\\
\white{.}\\
triplet excitation
\end{minipage} &
\begin{minipage}[c]{13cm}\includegraphics[width=12cm]{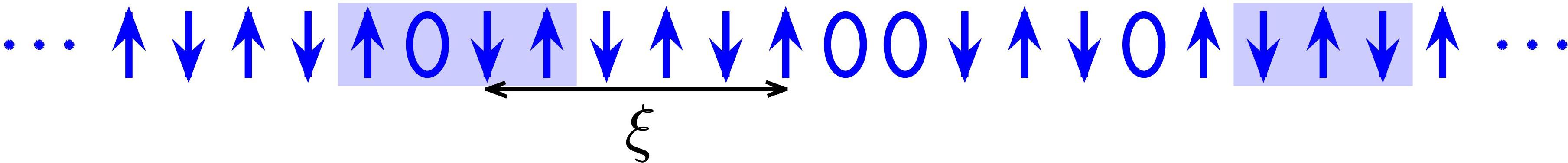} \\\includegraphics[width=12cm]{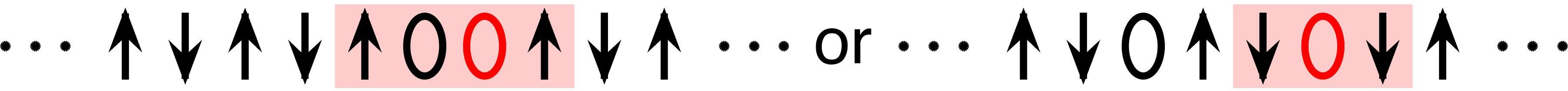} 
\end{minipage}
\end{tabular}
\end{center}

The gapped {\it Haldane triplet} excitation ruptures the string order in a topological (domain-wall like) manner.

\bigskip

An equivalent picture for the string order is achieved by decomposing each spin 1 into two spins 1/2 \cite{6}. On each site the spins 1/2 are paired symmetrically to a spin 1, between neighbouring sites they form a singlet pair state (antisymmetric pairing, {\it valence bond}).  This {\it valence bond solid} reflects best the unbroken translation symmetry of the Haldane ground state singlet.

\begin{center}
\includegraphics[width=\textwidth]{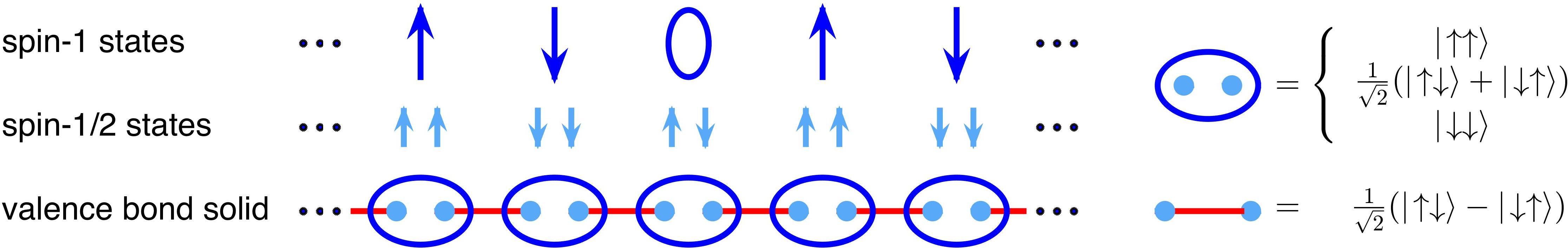}
\end{center}
\end{document}